\documentclass[english]{article}
\usepackage{lmodern}

\usepackage[T1]{fontenc}
\usepackage[latin9]{inputenc}
\usepackage{booktabs}
\usepackage{mathrsfs}
\usepackage{amsmath}
\usepackage{amssymb}
\usepackage{graphicx}

\makeatletter

\providecommand{\tabularnewline}{\\}

\newcommand{\lyxaddress}[1]{
	\par {\raggedright #1
	\vspace{1.4em}
	\noindent\par}
}

\usepackage{mathrsfs}   
\usepackage{bbold} 
\usepackage{url}
\usepackage{graphicx}
\usepackage[colorlinks=true,linkcolor=redLinks,citecolor=greenLinks,urlcolor=redLinks, pdfborder={0 0 1}]{hyperref}
\usepackage{xcolor}
\usepackage{framed}
\usepackage[numbers,sort&compress]{natbib}
\usepackage{amsmath}

\allowdisplaybreaks

\colorlet{shadecolor}{gray!15}

\definecolor{greenLinks}{rgb}{0, 0.6, 0} 
\definecolor{blueLinks}{rgb}{0, 0, 0.6}
\definecolor{redLinks}{rgb}{0.6, 0, 0}
\definecolor{tempText}{rgb}{0.55, 0.10,0.67}
\definecolor{eprintLinks}{rgb}{0.4, 0.4, 0.4}
\definecolor{journalLinks}{rgb}{0.6, 0, 0}

\newcommand{\MYhref}[3][redLinks]{\href{#2}{\color{#1}{#3}}}%

\usepackage{multirow}
\let\orig@Hy@EveryPageAnchor\Hy@EveryPageAnchor
\def\Hy@EveryPageAnchor{%
    \begingroup
    \hypersetup{pdfview=Fit}%
    \orig@Hy@EveryPageAnchor
    \endgroup
}

\let\oldFootnote\footnote
\newcommand\nextToken\relax

\renewcommand\footnote[1]{%
    \oldFootnote{#1}\futurelet\nextToken\isFootnote}

\newcommand\isFootnote{%
    \ifx\footnote\nextToken\textsuperscript{,}\fi}

\definecolor{myPurple}{RGB}{128,0,182}

\makeatother

\usepackage{babel}
\begin{document}
\title{Violation of lepton number in 3 units}
\author{Renato M. Fonseca\thanks{E-mail: fonseca@ipnp.mff.cuni.cz} \thanks{Contribution to the proceedings of the 6th Symposium on Prospects in the Physics of Discrete Symmetries (DISCRETE 2018), Vienna, Austria, 26--30 November 2018. This work is based on the paper \cite{Fonseca:2018ehk} which was written together with Martin Hirsch and Rahul Srivastava.} 
\date{}}
\maketitle

\lyxaddress{\begin{center}
{\Large{}\vspace{-0.5cm}}Institute of Particle and Nuclear Physics\\
Faculty of Mathematics and Physics, Charles University,\\
V Hole\v{s}ovi\v{c}k\'{a}ch 2, 18000 Prague 8, Czech Republic
\par\end{center}}
\begin{abstract}
The number of leptons may or may not be a conserved quantity. The
Standard Model predicts that it is (in perturbative processes), but
there is the well known possibility that new physics violates lepton
number in one or two units. The first case ($\Delta L=1$) is associated
to proton decay into mesons plus a lepton or an anti-lepton, while
the second one ($\Delta L=2$) is usually associated to Majorana neutrino
masses and neutrinoless double beta decay. It is also conceivable
that leptons can only be created or destroyed in groups of three ($\Delta L=3$).
Colliders and proton decay experiments can explore this possibility.
\end{abstract}

\section{Lepton number violation: looking beyond the usual scenarios}

There is currently no experimental hint of physical processes where
the total number of leptons (minus anti-leptons) changes, such as
\begin{align}
p & \rightarrow\pi^{0}e^{+}\;\textrm{ and }\;nn\rightarrow ppe^{-}e^{-}\,.\label{eq:normal-proton-decay}
\end{align}
Indeed, there are stringent bounds on the proton decay lifetime \cite{Miura:2016krn}
as well as on the neutrinoless double beta decay lifetime of several
isotopes \cite{KamLAND-Zen:2016pfg,Agostini:2018tnm}. Note that these
two processes are very different: the first destroys a baryon and
creates an anti-lepton ($\Delta B=\Delta L=-1$) so $B-L$ is preserved,
while the second creates two leptons ($\Delta L=2$) without changing
the number of baryons ($\Delta B=0$). Therefore it is possible that
one of these processes occurs and the other does not: if $B-L$ is
a conserved quantity then the proton may decay as indicated in (\ref{eq:normal-proton-decay})
but neutrinoless double beta decay is forbidden, while $B$ conservation
implies the converse.

These $\Delta L=\pm1,\pm2$ signals, as well as Majorana neutrino
masses and the production of pairs of same-sign leptons at colliders,
have been extensively studied over the years. But what if leptons
can only be created or destroyed in groups of three ($\Delta L=3$)?
 In this case all the processes mentioned so far would never be seen,
but the proton could still decay into three leptons,
\begin{equation}
p\rightarrow\ell^{\pm}\ell^{\pm}\ell^{\pm}+\textrm{mesons}\,,\label{eq:p-to-3-antileptons}
\end{equation}
and in a proton-proton collider it would be possible to produce three
same-sign leptons:
\begin{equation}
pp\rightarrow\ell^{\pm}\ell^{\pm}\ell^{\pm}+\textrm{jets}\,.
\end{equation}

One could object to the possibility of simultaneously observing lepton
number violation in three units at colliders and at proton decay experiments
with the argument that the current bounds on the proton's lifetime
rule out mediators of these processes with masses significantly below
$10^{16}$ GeV, hence these particles would not be produced at colliders.
However, this large mass bound applies only to proton decay channels
of the type shown in (\ref{eq:normal-proton-decay}) which are induced
by an effective 4-fermion interaction. The coefficient $c$ of such
a dimension 6 operator is proportional to the inverse of the square
of some new physics scale $\Lambda$,
\begin{equation}
c\sim\frac{1}{\Lambda^{2}}\,,\label{eq:c-coefficient}
\end{equation}
and in turn the proton decay width $\Gamma$ must be proportional
to the square of this quantity. We have to insert powers of the only
other energy scale in this process, the proton mass $m_{p}$,\footnote{This statement is not entirely true since there is also the QCD scale
$\Lambda_{QCD}\sim0.2$ GeV. However, given that $\Lambda_{QCD}$
is of the same order of magnitude as the proton mass, I will ignore
it.} in order to have an expression for $\Gamma$ with dimensions of energy,
so
\begin{equation}
\tau\left(\textrm{proton}\right)^{-1}=\Gamma\sim\frac{m_{p}^{5}}{\Lambda^{4}}\sim10^{32}\left(\frac{m_{p}}{\Lambda}\right)^{4}\textrm{years}^{-1}\,.
\end{equation}
Super-Kamiokande established that $\tau\left(p\rightarrow e^{+}\pi^{0}\right)>1.6\times10^{34}$
years at 90\% confidence level \cite{Miura:2016krn}, so from this
back-of-the-envelop calculation, it is clear why the mediator of this
process must have a large mass $\Lambda\gtrsim10^{16}$ GeV. However,
the effective interaction inducing the decay of a proton into three
charged leptons (or anti-leptons) has dimension 13 at the very least,
so instead of expression (\ref{eq:c-coefficient}) we have $c\sim\Lambda^{-9}$,
and the correct estimate for the proton's decay width as a function
of $\Lambda$ is
\begin{equation}
\tau\left(\textrm{proton}\right)^{-1}\sim10^{32}\left(\frac{m_{p}}{\Lambda}\right)^{18}\textrm{years}^{-1}\,.
\end{equation}
This formula shows that even for values of $\Lambda$ as low as a
few TeVs, the proton might still be sufficiently stable. It is worth
noting as well that at least 5 tracks will be produced in a Cherenkov
detector whenever a proton decays into 3 same-sign charged leptons,
which means that this signal should stand out cleanly over a very
low background. Nonetheless, no publicly available bounds on this
type of events seems to exist, in which case one must rely on very
old inclusive searches: $\tau\left(p\rightarrow e^{+}+\textrm{anything}\right)>0.6\times10^{30}$
years and $\tau\left(p\rightarrow\mu^{+}+\textrm{anything}\right)>1.2\times10^{31}$
years \cite{Reines:1974pb,Learned:1979gp}.

\section{The effective operator point of view}

In order to make model-independent statements about lepton number
violation in 3 units, it is convenient to look at non-renormalizable
interactions of the Standard Model (SM) fields with this property.
It is well known that all the renormalizable ones preserve baryon
and lepton number. Interestingly, non-perturbative effects associated
to an energy $E_{sph}$ of around 9 TeV violate both $L$ and $B$
in 3 units \cite{tHooft:1976rip,Klinkhamer:1984di}. It would certainly
be fascinating to produce for the first time in a laboratory an excess
of matter over anti-matter. But it is not certain that these processes
would be observable even if a parton center-of-mass energy $E_{sph}$
is reached in proton-proton collisions in the not-so-distant future
(see for example \cite{Ellis:2016ast,Tye:2017hfv,Cerdeno:2018dqk}).

Having said this, let us now go through non-renormalizable interactions
of Standard Model fields which violate $L$ and $B$. By convention
SM leptons have $U(1)_{L}$ charges of $\pm1$, so obvious any given
operator can only violate $L$ by an integer quantity. And the same
is true for baryon number, even though quarks carry a $\pm\frac{1}{3}$
charge under $U(1)_{B}$ --- one way to see this is by realizing
that color invariance requires an excess/deficit of quarks over anti-quarks
in each operator which must be a multiple of 3. In addition to this,
Lorentz invariance requires $\Delta B-\Delta L$ to be an even number.\footnote{The author of \cite{Kobach:2016ami} goes further, showing that $\frac{\Delta B-\Delta L}{2}+d$
is always even ($d$ is the operator dimension).} For example, this implies that if the proton is unstable, it must
decay into an odd number leptons. The lowest dimensional operators
which can induce this process are associated to $\Delta B=\Delta L=\pm1$:\footnote{Spinor and gauge indices will be suppressed throughout the text. The
fermion fields $q$, $u^{c}$, $d^{c}$, $l$ and $e^{c}$ are all
left-handed, which means that their conjugation (indicated with a
`{*}') produces a right-handed field.}
\begin{equation}
qqql,\;\;q\left(u^{c}\right)^{*}\left(d^{c}\right)^{*}l,\;\;qq\left(u^{c}\right)^{*}\left(e^{c}\right)^{*}\;\;\textrm{or }\left(u^{c}\right)^{*}\left(u^{c}\right)^{*}\left(d^{c}\right)^{*}\left(e^{c}\right)^{*}\,.
\end{equation}

As for $\Delta L=\pm3$, the simplest operators are expected to be
associated with $\Delta B=1$,\footnote{Without loss of generality, I will henceforth assume that $\Delta B$
is positive.} in which case their dimension $d$ cannot be smaller than 
\begin{equation}
\frac{3}{2}\left(n_{\textrm{leptons}}+n_{\textrm{quarks}}\right)\geq\frac{3}{2}\left(\left|\Delta L\right|+3\left|\Delta B\right|\right)=9\,.
\end{equation}
In fact there are two operators at the lower limit of this bound \cite{Weinberg:1980bf}:
\begin{equation}
\left(u^{c}\right)^{*}\left(u^{c}\right)^{*}\left(u^{c}\right)^{*}\left(e^{c}\right)^{*}ll\;\;\textrm{and }\left(u^{c}\right)^{*}\left(u^{c}\right)^{*}qlll\,.\label{eq:d=00003D9}
\end{equation}
However, in both cases the right-handed up quarks $u^{c}$ cannot
all be from the first generation (otherwise the operators become identically
zero), hence they do not induce proton decay. On the other hand, the
unique $d=10$ operator
\begin{equation}
\left(d^{c}\right)^{*}\left(d^{c}\right)^{*}\left(d^{c}\right)^{*}l^{*}l^{*}l^{*}h^{*}\label{e:d=00003D10}
\end{equation}
can induce the decay $p\rightarrow e^{-}\nu\nu\pi^{+}\pi^{+}$ ($\Delta L=-3$).
Note however that electric charge conservation implies that there
will necessarily be two neutrinos plus a charged lepton in the final state. It
is also worth point out that, from a theoretical point of view, it
is possible to forbid this operator with some discrete or continuous
symmetry: for example, if the laws of physics are $U\left(1\right)_{3B-L}$
symmetric, proton decay would be possible, but only through operators
of dimension 11, 13 or higher.

At $d=11$ one finds many $\Delta L=3$ operators with different structure,
for example
\begin{equation}
\partial\partial\left(u^{c}\right)^{*}\left(u^{c}\right)^{*}qlll\;\;\textrm{ and }\,\,\partial q\left(u^{c}\right)^{*}\left(u^{c}\right)^{*}ll\left(e^{c}\right)^{*}h\,,
\end{equation}
where the derivatives must be covariantly applied to the fields (or
alternatively a pair of derivatives can stand for a field strength
tensor). The problem with these operators is that experimentally they also
cannot be used to establish lepton number violation, since they involve
neutrinos. They induce processes such as $p\rightarrow e^{+}\overline{\nu}\overline{\nu}$
and $p\rightarrow\pi^{-}e^{+}e^{+}\overline{\nu}$.

If we insist on observing three charged leptons with the same sign,
then in order to conserve electric charge at least 5 quarks are needed
in the operator as well. That is an 8-fermion, dimension 12 operator,
but in reality it is easy to check that under the full Standard Model
group, there must also be either a derivative or a Higgs field in
the interaction, raising the operator dimension to 13. Two (out of
many) possibilities are
\begin{equation}
\partial\left(u^{c}\right)^{*}\left(u^{c}\right)^{*}\left(u^{c}\right)^{*}\left(u^{c}\right)^{*}d^{c}\left(e^{c}\right)^{*}\left(e^{c}\right)^{*}\left(e^{c}\right)^{*}\;\;\textrm{ and }\,\,\partial\left(u^{c}\right)^{*}\left(u^{c}\right)^{*}d^{c}qq\left(e^{c}\right)^{*}ll,
\end{equation}
both inducing the decay $p\rightarrow\ell^{+}\ell^{+}\ell^{+}\pi^{-}\pi^{-}$.

\section{A specific model}

The paper \cite{Fonseca:2018ehk} presents a specific model where
lepton number is violated in multiples of 3 units only. It contains
some fields that are not in the Standard Model (see table \ref{tab:new-fields}),
including some scalars which induce the normal $\Delta L=\pm1$ proton
decay modes under normal circumstances. However, a $Z_{3}\left(L\right)$
symmetry
\begin{equation}
\psi\to\omega^{L\left(\psi\right)}\psi
\end{equation}
with $\omega=\exp\left(2\pi i/3\right)$ will forbid the combinations
of interactions leading to such decays. For example, the scalar $s_{d}$
has gauge quantum numbers both of a leptoquark and of a diquark since
it couples to $ql$ and $q^{*}q^{*}$. Therefore, it would generate
the effective operator $qqql$ were it not for the $Z_{3}\left(L\right)$
symmetry which forbids the diquark coupling.

\begin{figure}[tbph]
	\begin{centering}
		\includegraphics[scale=0.5]{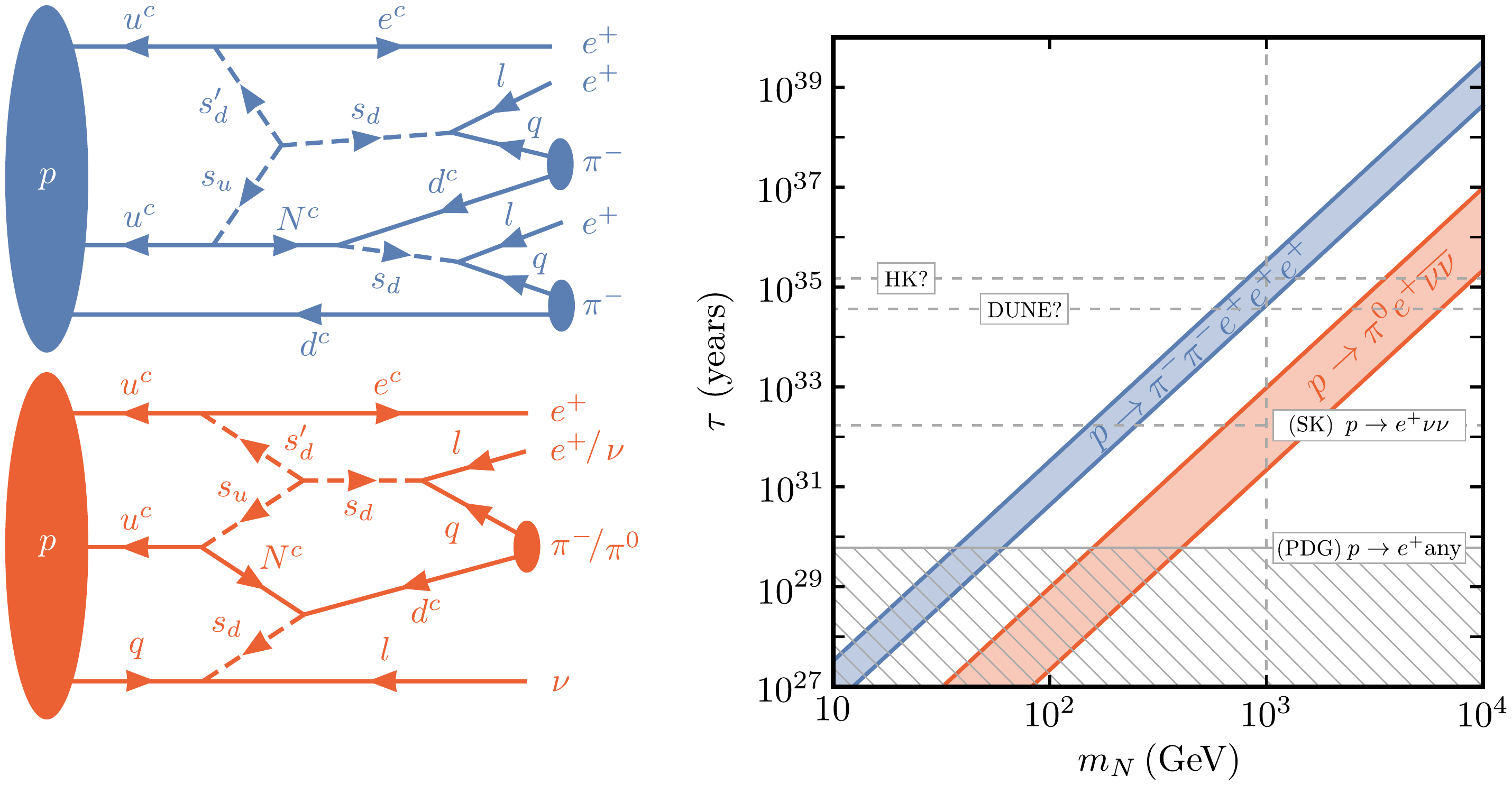}
		\par\end{centering}
	\caption{\label{fig:1}On the left: diagrams of the two main contributions
		to proton decay in the model described in the main text. On the right:
		rough proton lifetime estimates as a function of the mass $m_{N}$,
		for 1 TeV scalar masses and order 1 Yukawa couplings. This figure
		was adapted from \cite{Fonseca:2018ehk}.}
\end{figure}

\begin{table}[tbph]
	\begin{centering}
		\begin{tabular}{cccc}
			\toprule 
			Field(s) & Spin & $SU(3)_{C}\times SU(2)_{L}\times U(1)_{Y}$ & $L$\tabularnewline
			\midrule
			$N$ $(\times3)$ & Left-fermion & $\left(\boldsymbol{1},\boldsymbol{1},0\right)$ & $+1$\tabularnewline
			$N^{c}$ $(\times6)$ & Left-fermion & $\left(\boldsymbol{1},\boldsymbol{1},0\right)$ & $-1$\tabularnewline
			$s_{u}$ & Scalar & $\left(\overline{\boldsymbol{3}},\boldsymbol{1},-\frac{2}{3}\right)$ & $-1$\tabularnewline
			$s_{d},s_{d}^{\prime}$ & Scalar & $\left(\overline{\boldsymbol{3}},\boldsymbol{1},\frac{1}{3}\right)$ & $-1$\tabularnewline
			\bottomrule
		\end{tabular}
		\par\end{centering}
	\caption{\label{tab:new-fields}New fields in the model presented in \cite{Fonseca:2018ehk}.}
\end{table}
The interactions of the new fields allowed by all symmetries are
\begin{align}
\mathscr{L} & =Y_{\nu}lN^{c}h+Y_{1}\left(u^{c}\right)^{*}\left(N^{c}\right)^{*}s_{u}+Y_{2}N^{c}d^{c}s_{d}^{*}+Y_{3}\left(e^{c}\right)^{*}\left(u^{c}\right)^{*}s_{d}^{\prime}+Y_{4}qls_{d}\nonumber \\
& +\mu s_{u}s_{d}s_{d}^{\prime}+m_{N}NN^{c}\,.\label{eq:Lagrangian}
\end{align}
There are several important remarks to be made about this Lagrangian:
\begin{enumerate}
	\item In the absence of the $\mu$ term, lepton and baryon number are conserved.
	Therefore the scalar trilinear interaction $s_{u}s_{d}s_{d}^{\prime}$
	must appear in any diagram with a net number of external baryons/leptons
	different from zero.
	\item By construction the model is $Z_{3}\left(L\right)$ symmetric, but
	it turns out that the Lagrangian is invariant under the bigger $U\left(1\right)_{3B-L}$
	symmetry group. This is significant because the $Z_{3}\left(L\right)$
	symmetry would allow in principle $\Delta\left(B,L\right)=\left(1,-3\right)$
	processes, while the $U\left(1\right)_{3B-L}$ symmetry does not.
	Hence, in accordance with what was discussed previously, proton decay
	is induced only by operators of dimension 11, 13, or higher.
	\item Majorana masses are forbidden so, as expected, neutrinos are Dirac
	particles. In this model the three small masses of the neutrinos observed
	in oscillation experiments can be obtained through a rather delicate
	choice of the matrices $Y_{\nu}$ and $m_{N}$.
	\item The fields $s_{d}$ and $s_{d}^{\prime}$ have the same quantum numbers,
	therefore they can have the same interactions. However, meson decay
	and atomic parity violation experiments place stringent bounds on
	the product of some couplings \cite{Davidson:1993qk,Dorsner:2016wpm}.
	To avoid them, one might consider that $s_{d}$ interacts only with
	left-handed Standard Model fermions, while $s_{d}^{\prime}$ couples
	exclusively with right-handed ones. The Lagrangian in equation (\ref{eq:Lagrangian})
	assumes so.
\end{enumerate}
The most important diagrams for proton decay are shown in figure \ref{fig:1}.
An interesting way to see that the proton can only decay through these
complicated diagrams is to start with the crucial trilinear interaction
$s_{u}s_{d}s_{d}^{\prime}$ and connect each of these scalars to Standard
Model fermions through the available Yukawa interactions. Using some
optimistic values for the various parameters of the model (1 TeV scalar
masses, and order 1 Yukawa couplings) and leaving free the heavy neutrino
mass $m_{N}$ one obtains estimates for the partial lifetimes $\tau\left(p\to e^{+}e^{+}e^{+}\pi^{-}\pi^{-}\right)$
and $\tau\left(p\to\pi^{0}e^{+}\overline{\nu}\overline{\nu}\right)$
which might be within reach of future proton decay experiments. It
depends on $m_{N}$, as shown in figure \ref{fig:1}.

Lepton number violation can be observed at the LHC through the pair
production of one of the scalar leptoquarks ($s_{u}$, $s_{d}$ or
$s_{d}^{\prime}$), followed by their decay into two different final
states. In order to see several of these events, each of the two decay
modes should have a large branching fraction. The leptoquarks $s_{d}^{\left(\prime\right)}$
do not fulfill this requirement because they yield either a 2-body
or a 6-body final state, hence the branching fraction of the second
decay mode is very small. On the other hand, $s_{u}$ has two distinct
4-body final states: $s_{u}\rightarrow s_{d}^{*}+s_{d}^{\prime*}\rightarrow2\ell^{-}+2\textrm{jets}$
and $s_{u}\rightarrow N^{c}+u^{c}\rightarrow\ell^{+}+3\textrm{jets}$.
The branching fractions are similar if $s_{d}$, $s_{d}^{\prime}$
and $N^{c}$ are produced off-shell. Overall, one would see three
same-sign leptons in a proton-proton collision: $pp\to s_{u}s_{u}^{*}\to3\ell^{\pm}+\textrm{jets}$.

\section*{Summary}

Lepton number violation is almost always associated with processes
where one or two leptons are created or destroyed. Well known examples
are proton decay into a (anti)lepton plus a meson, and neutrinoless
double beta decay. However, it is conceivable that lepton number can
only be changed in multiples of 3, in which case the signals to look
for are different. In fact, this is what happens in the Standard Model
since non-perturbative effects can change the net lepton and baryon
numbers by 3 units only.

Following \cite{Fonseca:2018ehk}, I presented here the possibility
that there might be perturbative new physics at the TeV scale associated
to $\Delta L=3$ and $\Delta B=1$. Even though the new fields responsible
for lepton and baryon number violation are much lighter than usually
assumed, a simple estimate indicates that the proton decay lifetime
can easily excess the current experimental bounds. The reason is simple:
$U\left(1\right)_{3B-L}$ invariance implies that proton decay is
induced by high dimensional operators only, which means that the proton
lifetime is enhanced by many powers of the ratio of the new physics
scale over the proton mass. Hyper-Kamiokande and DUNE might be able
to observe the decay $p\rightarrow e^{+}e^{+}e^{+}\pi^{-}\pi^{-}$,
and at the LHC it might be possible to produce 3 leptons with the
same charge plus jets and no missing energy.

\section*{Acknowledgments}

I am thankful to the two other authors of \cite{Fonseca:2018ehk},
Martin Hirsch and Rahul Srivastava, for the fruitful collaboration.
I would also like to acknowledge the financial support from the Grant
Agency of the Czech Republic (GA\v{C}R) through contract number 17-04902S,
as well as from the Generalitat Valenciana through the grant SEJI/2018/033.


\begin{thebibliography}{10}
\providecommand{\url}[1]{\texttt{#1}}
\providecommand{\urlprefix}{URL }
\providecommand{\eprint}[2][]{\url{#2}}

\bibitem{Fonseca:2018ehk}
R.~M. Fonseca, M.~Hirsch and R.~Srivastava, \emph{{$\Delta L = 3$ processes:
  Proton decay and the LHC}},
  \MYhref[journalLinks]{http://dx.doi.org/10.1103/PhysRevD.97.075026}{Phys.
  Rev.
  }\MYhref[journalLinks]{http://dx.doi.org/10.1103/PhysRevD.97.075026}{\textbf{D97}
  (2018) 7 075026},
  \MYhref[eprintLinks]{http://arxiv.org/abs/1802.04814}{{\ttfamily
  arXiv:1802.04814 [hep-ph]}}.

\bibitem{Miura:2016krn}
K.~Abe et~al. (Super-Kamiokande), \emph{{Search for proton decay via $p \to
  e^+\pi^0$ and $p \to \mu^+\pi^0$ in 0.31 megaton$\cdot$years exposure of the
  Super-Kamiokande water Cherenkov detector}},
  \MYhref[journalLinks]{http://dx.doi.org/10.1103/PhysRevD.95.012004}{Phys.
  Rev.
  }\MYhref[journalLinks]{http://dx.doi.org/10.1103/PhysRevD.95.012004}{\textbf{D95}
  (2017) 1 012004},
  \MYhref[eprintLinks]{http://arxiv.org/abs/1610.03597}{{\ttfamily
  arXiv:1610.03597 [hep-ex]}}.

\bibitem{KamLAND-Zen:2016pfg}
A.~Gando et~al. (KamLAND-Zen), \emph{{Search for Majorana neutrinos near the
  inverted mass hierarchy region with KamLAND-Zen}},
  \MYhref[journalLinks]{http://dx.doi.org/10.1103/PhysRevLett.117.082503}{Phys. Rev. Lett.
  }\MYhref[journalLinks]{http://dx.doi.org/10.1103/PhysRevLett.117.082503}{\textbf{117} (2016) 8 082503}, [Addendum:
  \MYhref[journalLinks]{http://dx.doi.org/10.1103/PhysRevLett.117.109903}{Phys. Rev. Lett. \textbf{117} (2016) 10 109903}],
  \MYhref[eprintLinks]{http://arxiv.org/abs/1605.02889}{{\ttfamily
  arXiv:1605.02889 [hep-ex]}}.

\bibitem{Agostini:2018tnm}
M.~Agostini et~al. (GERDA), \emph{{Improved limit on neutrinoless
  double-$\beta$ decay of $^{76}$Ge from GERDA phase II}},
  \MYhref[journalLinks]{http://dx.doi.org/10.1103/PhysRevLett.120.132503}{Phys.
  Rev. Lett.
  }\MYhref[journalLinks]{http://dx.doi.org/10.1103/PhysRevLett.120.132503}{\textbf{120}
  (2018) 13 132503},
  \MYhref[eprintLinks]{http://arxiv.org/abs/1803.11100}{{\ttfamily
  arXiv:1803.11100 [nucl-ex]}}.

\bibitem{Reines:1974pb}
F.~Reines and M.~F. Crouch, \emph{{Baryon conservation limit}},
  \MYhref[journalLinks]{http://dx.doi.org/10.1103/PhysRevLett.32.493}{Phys.
  Rev. Lett.
  }\MYhref[journalLinks]{http://dx.doi.org/10.1103/PhysRevLett.32.493}{\textbf{32}
  (1974) 493--494}.

\bibitem{Learned:1979gp}
J.~Learned, F.~Reines and A.~Soni, \emph{{Limits on nonconservation of baryon
  number}}, \MYhref[journalLinks]{http://dx.doi.org/10.1103/PhysRevLett.43.907}{Phys. Rev. Lett.
  }\MYhref[journalLinks]{http://dx.doi.org/10.1103/PhysRevLett.43.907}{\textbf{43} (1979) 907}, [Erratum: \MYhref[journalLinks]{http://dx.doi.org/10.1103/PhysRevLett.43.1626}
  {Phys. Rev. Lett. \textbf{43} (1979) 1626}].

\bibitem{tHooft:1976rip}
G.~'t~Hooft, \emph{{Symmetry breaking through Bell-Jackiw anomalies}},
  \MYhref[journalLinks]{http://dx.doi.org/10.1103/PhysRevLett.37.8}{Phys. Rev.
  Lett.
  }\MYhref[journalLinks]{http://dx.doi.org/10.1103/PhysRevLett.37.8}{\textbf{37}
  (1976) 8--11}, [,226(1976)].

\bibitem{Klinkhamer:1984di}
F.~R. Klinkhamer and N.~S. Manton, \emph{{A saddle point solution in the
  Weinberg-Salam theory}},
  \MYhref[journalLinks]{http://dx.doi.org/10.1103/PhysRevD.30.2212}{Phys. Rev.
  }\MYhref[journalLinks]{http://dx.doi.org/10.1103/PhysRevD.30.2212}{\textbf{D30}
  (1984) 2212}.

\bibitem{Ellis:2016ast}
J.~Ellis and K.~Sakurai, \emph{{Search for sphalerons in proton-proton
  collisions}},
  \MYhref[journalLinks]{http://dx.doi.org/10.1007/JHEP04(2016)086}{JHEP
  }\MYhref[journalLinks]{http://dx.doi.org/10.1007/JHEP04(2016)086}{\textbf{04}
  (2016) 086}, \MYhref[eprintLinks]{http://arxiv.org/abs/1601.03654}{{\ttfamily
  arXiv:1601.03654 [hep-ph]}}.

\bibitem{Tye:2017hfv}
S.~H.~H. Tye and S.~S.~C. Wong, \emph{{Baryon number violating scatterings in
  laboratories}},
  \MYhref[journalLinks]{http://dx.doi.org/10.1103/PhysRevD.96.093004}{Phys.
  Rev.
  }\MYhref[journalLinks]{http://dx.doi.org/10.1103/PhysRevD.96.093004}{\textbf{D96}
  (2017) 9 093004},
  \MYhref[eprintLinks]{http://arxiv.org/abs/1710.07223}{{\ttfamily
  arXiv:1710.07223 [hep-ph]}}.

\bibitem{Cerdeno:2018dqk}
D.~G. Cerde\~no, P.~Reimitz, K.~Sakurai and C.~Tamarit, \emph{{$B+L$ violation
  at colliders and new physics}},
  \MYhref[journalLinks]{http://dx.doi.org/10.1007/JHEP04(2018)076}{JHEP
  }\MYhref[journalLinks]{http://dx.doi.org/10.1007/JHEP04(2018)076}{\textbf{04}
  (2018) 076}, \MYhref[eprintLinks]{http://arxiv.org/abs/1801.03492}{{\ttfamily
  arXiv:1801.03492 [hep-ph]}}.

\bibitem{Kobach:2016ami}
A.~Kobach, \emph{{Baryon number, lepton number, and operator dimension in the
  Standard Model}},
  \MYhref[journalLinks]{http://dx.doi.org/10.1016/j.physletb.2016.05.050}{Phys.
  Lett.
  }\MYhref[journalLinks]{http://dx.doi.org/10.1016/j.physletb.2016.05.050}{\textbf{B758}
  (2016) 455--457},
  \MYhref[eprintLinks]{http://arxiv.org/abs/1604.05726}{{\ttfamily
  arXiv:1604.05726 [hep-ph]}}.

\bibitem{Weinberg:1980bf}
S.~Weinberg, \emph{{Varieties of baryon and lepton nonconservation}},
  \MYhref[journalLinks]{http://dx.doi.org/10.1103/PhysRevD.22.1694}{Phys. Rev.
  }\MYhref[journalLinks]{http://dx.doi.org/10.1103/PhysRevD.22.1694}{\textbf{D22}
  (1980) 1694}.

\bibitem{Davidson:1993qk}
S.~Davidson, D.~C. Bailey and B.~A. Campbell, \emph{{Model independent
  constraints on leptoquarks from rare processes}},
  \MYhref[journalLinks]{http://dx.doi.org/10.1007/BF01552629}{Z. Phys.
  }\MYhref[journalLinks]{http://dx.doi.org/10.1007/BF01552629}{\textbf{C61}
  (1994) 613--644},
  \MYhref[eprintLinks]{http://arxiv.org/abs/hep-ph/9309310}{{\ttfamily
  arXiv:hep-ph/9309310 [hep-ph]}}.

\bibitem{Dorsner:2016wpm}
I.~Dor\v{s}ner et~al., \emph{{Physics of leptoquarks in precision experiments
  and at particle colliders}},
  \MYhref[journalLinks]{http://dx.doi.org/10.1016/j.physrep.2016.06.001}{Phys.
  Rept.
  }\MYhref[journalLinks]{http://dx.doi.org/10.1016/j.physrep.2016.06.001}{\textbf{641}
  (2016) 1--68},
  \MYhref[eprintLinks]{http://arxiv.org/abs/1603.04993}{{\ttfamily
  arXiv:1603.04993 [hep-ph]}}.

\end{thebibliography}
\end{document}